# A new model for Context-Oriented Programs

Mohamed A. El-Zawawy[1,2], Eisa A. Aleisa[1]
[1]College of Computer and Information Sciences, Al Imam Mohammad Ibn Saud Islamic University (IMSIU), Riyadh, Kingdom of Saudi Arabia
[2]Department of Mathematics, Faculty of Science, Cairo University, 12613, Giza, Egypt
maelzawawy@ccis.imamu.edu.sa

**Abstract:** Context-oriented programming (COP) is a new technique for programming that allows changing the context in which commands execute as a program executes. Compared to object-oriented programming (aspect-oriented programming), COP is more flexible (modular and structured). This paper presents a precise syntax-directed operational semantics for context-oriented programming with layers, as realized by COP languages like ContextJ* and ContextL. Our language model is built on Java enriched with layer concepts and activation and deactivation of layer scopes. The paper also presents a static type system that guarantees that typed programs do not get stuck. Using the means of the proposed semantics, the mathematical correctness of the type system is presented in the paper.

## 1. Introduction

Modularity of performance alterations relies on the dynamic environment of program executions. Context-oriented programming (COP) (Hirschfeld, Costanza & Nierstrasz, 2008) emerged as a programming technique to enhance this modularity. Classically these performance alterations are distributed among program modules and usually complex engineering is necessary to back dynamic combination of the modules. Smalltalk (Golubski & Lippe, 1995), Java (Campione, Walrath & Huml, 2000), JavaScript (Flanagan, 2012), and Common Lisp (Costanza, Herzeel & D'Hondt, 2009) are examples of languages on which COP were established. The base languages for COP are typical object oriented languages. Main features of COP include (a) layers of variant procedures for introducing and classifying performance alterations and (b) an instrument for layer activation to endorsement and composition. A variant procedure is a procedure that can be executed around, after, or before the same (variant) procedure defined in a different part (class or layer) of the program. A layer is a set of variant procedures. A layer can be (de)activated in main function. Layers are meant to determine the specific semantics of objects for adaption with different applications.

In this paper, we present a new model for COP. The proposed model has basic language features. The model has the advantage of extending directly over well-studied Java features. The model is in-complex yet articulates enough to include more language features. Besides typical Java features, the model provides overriding (i.e., around-type) variant procedures, layers activation and deactivation, and a call mechanism for *proceed* and *super*. This paper also presents an operation semantics that directly (without mapping to non-COP) models the meanings of basic COP constructs. For the core of COP languages, the proposed semantics can be used to provide precise specifications. The paper also presents a type system for COP. Typically; a type system statically ensures the absence of run-time errors such as procedure-not-found and field-not-found errors. Noticeably, establishing the type system is not an easy task because in COP the existence of a procedure definition in a class may well rely upon whether a specific layer is activated. The paper also provides a mathematical proof for the soundness of the type system based on the proposed operational semantics.

## Example

Figure 1 provides a COP example. Class Cube defines three variables of type integer (*length, width*, and *height*) with a constructor for initialization. The class also includes the *modify()* procedure to modify different variables.

The first definition of *modify()* is the main one and modifies and shows *length*. This definition is included in the main layer which is effectual for all objects of Cube. The second definition of *modify()* is a refinement and is included in the layer *Second_dim*. This refinement modifies *width* and appends its new value (the second dimension of the cube) that might be needed for further calculations. This refinement is effective only when its layer is activated. The third definition of *modify()* is yet another refinement and is included in the layer *Third_dim*.

In the example of Figure 1, the refinements of *modify()* runs the command *proceed()*. This special command invokes all refinements of *modify()* included in layers already activated ahead of the activation of the *Second_dim* or *Third dim* layer. This command also invokes the version of *modify()* included in the main layer. On the other hand, the *super* command included in our language model (Figure 2) starts the lookup for procedures from the super-class of the class containing the current procedure.

```
Class Cube{
    int length, width, height;
    cube(int val1, int val2, int val2)
      { length:=val1; wedith:=val2; height:=val3; }
    modify()
       { length:= 4; return "Length:" + length; }
    layer Second_dim
       { modify()
          { width:= 5; return
proceed+ ";Width:  "+width;     }}
  layer Third_dim
      { modify()
        { height:= 6; return proceed+"; Height:"+height;
}}}
```

**Figure 1**: A COP program

The *with* and *without* constructs are used in COP for layer activation and deactivation, respectively. We show their use on the following object of the class Cube.

*Cube c(1, 1, 1);*

While no layers are activated, the following standard command invokes the version of the main layer of *modify()* that modifies and returns only the length of the cube c.

*System.out.println(c);*

*=> "Length: 4"*

However, the following example activates *Second_dim* layer (via *with*). In this case, the printing command invokes first the version of *modify()* included in *Second_dim* and then invokes the version of the main layer of *modify()*.

*with Second_dim (System.out.println(c));*

*=>"Length: 4; width: 5;"*

Another example is the following:

*without Second dim (with Second dim (System.out.println(c)));*

*=>"Length: 4;"*

**Contributions**

Contributions of this paper are the following:

1. A precise operational semantics for a rich model of context-oriented programming languages.
2. A static type system that is mathematically sound for context-oriented programming languages.

**Organization**

The organization of the rest of the paper is as follows. Section 2 presents the language model and the operational semantics of the language. The type system together with its mathematical soundness proof is presented in Section 3. Related and future work is discussed in Section 4.

## 2. Syntax and Operational Semantics

This section presents the model of our programming language together with an operational semantics for the language. Most basic object-oriented aspects as subtyping and inheritance are included in the language (dubbed J-COP) that we use in this paper. For the sake of readability, we followed the Java syntax for corresponding constructs. The syntax of J-COP is shown in Figure 2.

*Bool* and *int* are our primitive types. We assume that $\mathbb{C}$ is a set of class names with typical element $C$. The set of types (Types) includes *bool*, *int*, and $\mathbb{C}$. Moreover "Types" has reference and function types. We let $\tau$ be a typical element of the set of types. We let *LVar* denotes the set of local variables. Local variables are contained in procedures and are active as long as their hosting procedures are active. Local variables also serve as parameters for procedures. The set of instance variables of a class $C$ is denoted by $Var_C$. The internal state of a class is stored via its instance variables. Typical elements of *IVar* and $IVar_C$ are $o$ and $v$, respectively. The sets of procedure and layer names are denoted by *FunNames* (typical element is $f$) and *LayerNames* (typical element is $l$), respectively. A layer expression is a sequence of layer activation/deactivation. A typical element of the set of layer expressions, denoted by *LayerExpr*, is denoted by $le$.

A program in *J-COP* consists of a set of classes and a main procedure triggering the program execution. A class contains definitions for a set of procedures and a set of layers each of which contains the definition of a procedure. A parameter, a statement, and an expression are the components of a procedure where the expression denotes the value returned by the procedure.

We use a state representation and a subtype relation to define an operational semantics for the language *J-COP*. We let $\tau_1 \leq \tau_2$ denotes that $\tau_1$ is a subtype of $\tau_2$. The class definitions of a given program are used to build the relation $\leq$ which is introduced in Definition 1.

**Definition 1**

1. Types = $\{bool, int, C, ref\ \tau, \tau_1 \rightarrow \tau_2\}$.
2. A class $C$ is a subclass of a class $D$ (denoted by $C \ll D$) if $C$ inherits $D$ by definition of $C$. The relation $\leq_C$ on the set of classes is the reflexive transitive closure of $\ll$. A class $D$ is a superclass of $C$, if $C$ is a subclass of $D$.

3. The order $\leq$ on the set of types is defined as: $\leq_C \cup \{\tau \leq \tau \mid \tau \in \{\text{int}, \text{bool}, \text{ref}\,\tau, \tau_1 \rightarrow \tau_2\}\}$

$\tau \in Types ::= int \mid bool \mid C \mid ref\ \tau \mid \tau_1 \rightarrow \tau_2$
$e \in Exprs ::= n \mid (C)e \mid this \mid o \mid e.v \mid e_1\, i_{op}\, e_2$
$b \in Bexprs ::= true \mid false \mid e_1 c_{op}\, e_2 \mid b_1 b_{op} b_2$
$le \in LayerExprs ::= with\ l \mid \text{without}\ l \mid \epsilon \mid le\ le$
$S \in Stmts ::= e_1.v := e_2 \mid o_1 := le\ o_2.f(e) \mid o_1 := o_2.f(e) \mid o_1 := super.f(e)\ o_1 := proceed\ o_2.f(e) \mid o := new\ C \mid S_1; S_2 \mid if\ b\ then\ S_t else\ S_f \mid while\ b\ do\ S_t$
$fun \in Funs ::= f(p)\{S; return(e);\}$
$layer \in Layers ::= Layer\ l\ \{fun\}$
$inhrt \in Inherits ::= \epsilon \mid inherits\ C$
$class \in Classes: := class\ C\ inhrt\ \{fun^*\ layer^*\}$
$prog \in Progs ::= class^* main()\{S\}$

**Figure** 2: The programming language J-COP

Definition 2 introduces necessary components towards introducing the states of the operational semantics. The symbol $\mathcal{A}$ denotes an infinite set of memory addresses with α as a typical element of $\mathcal{A}$.

**Definition 2**

1. For a class $C$, $IVar_C$ and $Fun_C$ denote the set of instance variables and the set of functions of $C$, respectively. The set of layer names of a class $C$ is denoted by $Layer_C$.
2. $\wp = \mathbb{Z} \cup \mathcal{A} \cup \{\perp\}$.
3. $Stacks = \{s \mid s: LVar \rightarrow \wp\}$.
4. $ObjectContents = \{I_{(C,n)} \mid I_{(C,n)}: IVar_C \rightarrow \wp, C \in \mathbb{C}, n \in \mathbb{N}\}$.
5. $\text{Heaps} = \{\text{h} \mid \text{h}: \mathcal{A} \rightarrow_p \{(C, n, I_{(C,n)}) \mid C \in \mathbb{C}, n \in \mathbb{N}\}$.
6. $States = \{(s, h, L^s) \mid s \in stacks, h \in Heaps, L^s \subseteq LayerNames\}$.

Model values are elements of the set $\wp$. A semantic state is a triple of a stack, a heap, and a set of layer names that are active at that program point (state). The set of local variables includes the special variable *this* which points at the current active object. For an address $\alpha \in dom(h)$, $h_i(\alpha)$ denotes the $i^{th}$ component of the triple $h(\alpha)$, where $i = 1,2,3$.

Definition 3 introduces the notations $F_C$ and $L_C$. For a class $C$, $F_C$ maps each procedure name in $C$ to the triple consisting of the parameter variable of the procedure, procedure body, and returned expression of procedure. For a class $C$, $I_C$ maps each layer name in $C$ to the components of its procedure.

**Definition 3**

1. $FunBodies = \{F_C \mid F_C: Fun_C \rightarrow LVar \times Stmt \times Expr; f \mapsto (p_f, S_f, e_f)\}$.
2. $Layers = \{L_C \mid L_C: Layer_C \rightarrow Fun_C \times LVar \times Stmt \times Expr; f \mapsto (f, p_f, S_f, e_f)\}$.

Figure 3 presents inference rules of four procedures that are used in the inference rules of the operational semantics.

For a given list of layer names *Ls* and a layer expression *le*, Figure 3 presents the procedure *layer* which adds the layers activated by *le* to *Ls* and removes the layers deactivated by *le* from *Ls*. The definition of the class procedure is presented in Figure 3. This procedure finds whether a given variable belongs to a given class or to any of its ancestor classes. The procedure *super*, which for a function name and a class name searches for the first ancestor of the class that contains a definition for the function, is outlined in the same figure which as well presents the definition of the procedure *clslyrs*. This procedure determines which members of a given list of active layers (*L*) contain a definition for a given procedure, *f*.

The semantics of the *J-COP* expressions is presented in Figure 4. Some comments on the figure are in order. The variable $v$ of the class pointed-to by *e* is denoted by $e.v$. We assume that the set of variables in a class does not intersect with the set of the variables of any of the class's ancestors. We also assume that for a class $C$, the domain of $IVar_C$ includes all the variables of $C$ and its ancestors. Hence the rule $inst_1^s$ ensures that $v$ is a member of the class pointed-to by *e* or is a member of any of the class's ancestors (via calling the *class* procedure). The semantic of *e* is the address of the triple in memory representing the meant class object. The third component of this triple is denoted by *I* (which is a map representing the values of the object's variables). The rule $cast_1^s$ says that the cast of the expression *e* in the form of a class $C$ aborts only if *e* points to a triple in the memory that represents a class $D$ that is not a descendant of $C$.

Definition 4 formalizes the case when a statement aborts execution.

**Definition 4**

A statement $S$ aborts at a state $(s, h, L^s)$, denoted by $S: (s, h, L^s) \rightarrow abort$, if it not possible (provided that $S$ is not stuck in an infinite loop) to find a state $(s', h', L^{s\prime})$ such that $S: (s, h, L^s) \rightarrow (s', h', L^{s\prime})$ according to inference rules of Figure 5.

The semantics of the statements of the *J-COP* language is shown in Figure 5. Some comments on the rules are as follows. The rule $(:=_e^s)$ modifies the variable $v$ of the object referenced by $e_1$. This is done via updating the third component of $h([\![e_1]\!](s,h))$ and keeping the first two components $(h_2([[\![e_1]\!](s,h))$ and $h_3([\llbracket e_1 \rrbracket(s,h)]))$ of the triple unchanged. The semantics of executing the function $f$ of the object referenced by $o_2$ on the input $e$ is captured by the rule $(\coloneqq^s_{o.f})$. The expression $le$ of the statement $o_1 \coloneqq o_2.f(e)$ activates/deactivates some specific layers. The semantic of the statement is given via the rule $(\coloneqq^s_{l.o.f})$. This rule first adds (removes) activated (deactivated) layers of $le$ to $L^s$ to produce $L^s_1$ (via calling the procedure *layer*). The rule then finds the sub-list of $L^s_1$ whose elements contain a definition for the function $f$. Then the rule sequentially executes these functions. The execution of a function definition considers the previous execution via modifying $o_1$ to the $\llbracket e_{i-1} \rrbracket(s_i,h_i)$ . The rule $(sup^s)$ expresses the semantics of the statement $o_1 \coloneqq super.f(e)$ which executes the procedure $f$ defined in an ancestor of the current class. This ancestor that hosts $f$ is found using the procedure *super*. The rule $(pro^s)$ introduces the semantics of the statement $o_1 \coloneqq proceed\ o_2.f(e)$ which executes all functions named $f$ and contained in an active layer of the object pointed-to by $o_2$. The procedure *clylyrs* is used in this rule to decide which of the currently active layers $(L^s)$ contains a definition for $f$ and is a member of the object pointed-to by $o_2$.

## 3. Type System

This section presents a type system for the language *J-COP*. The function of the type system is to statically detect type errors like *variable-not-found* and *procedure-not-found*. Our type system also assures success of *proceed()* and *super()* calls. The concept of layer activation/deactivation makes developing such type system is not an easy task. This is so because layer activation/deactivation affects the list of procedures to be considered included in a given class. Definition 5 presents the context definition.

**Definition 5.** *1.* $Var = LVar \cup (\cup\{(C,v) \mid v \in IVar_c \text{ and } C \text{ is a class}\})$.

*2. The set of contexts is defined as* $\{(\Gamma, L^t) \mid \Gamma: Var \rightarrow_p \{int,\ ref\ C\} and\ L^t \subseteq LayerNames\}$.

The proposed type system for the *J-COP* language is shown in Figure 6. Some comments on the rules are as follows. For expressions, the type judgment has the form $\Gamma \vDash e{:}\tau$, read "$e$ is of type $\tau$ under $\Gamma$ where $\Gamma$ denotes a finite function from variables to the set *{int, ref C}*. For class procedures, the type judgment has the form $(\Gamma, L^t) \vDash (C,f){:}\tau_1 \rightarrow \tau_2$, read "the procedure $f$ of the class $C$ is of type $\tau_1 \rightarrow \tau_2$ under $\Gamma$ and $L^t$ " where $L^t$ denotes a set of active layers. For layer procedures, the type judgment has the form $(\Gamma, L^t) \vDash (C,f,l){:}\tau_1 \rightarrow \tau_2$ , read "the procedure f of the layer l contained in the class C is of type $\tau_1 \rightarrow \tau_2$ under $\Gamma$ and $L^t$". For statements, the type judgment has the form $(\Gamma, L^t) \vDash S{:}WF$, read "$S$ is well formed and safe to be executed under $C$ and $L^t$".

The precondition of the rule $(C.f^t)$ requires that the body $S$ of the procedure $f$ to be well formed. The precondition also requires the existence of a common type that covers any overloading for $f$. The first part of the precondition of the rule $(\coloneqq^t_{l.o.f})$ requires that all procedures named $f$ inside layers of the class $C$ to have an upper bound type. Among others requirements, the precondition of this rule also ensures that the set $L^t$ is in line with the expression $\leq (layer(le, L^t) = L^t)$. The rule $(pro^t)$ uses the rule $(\coloneqq^t_{l.o.f})$ to determine types for all instances of $f$ in layers of the class $C$. In line with expectation of the rules for non-atomic statements like $(if^t), (while^t),\ and\ (ser^t)$, these rules require their sub-statements to be well formed.

Definition 6 presents the condition when a state respects a context denoted by $(s,h,L^s) \sim (\Gamma, L^t)$.

**Definition 6.**

*1.* $(s,h,L^s) \sim (\Gamma, L^t) \Leftrightarrow^{def}$

*(a)* $L^s \subseteq L^t$,

*(b)* $\forall\, o \in dom(\Gamma).\Gamma(o) = int \Rightarrow s(a) \in \mathbb{Z}$ ,

*(c)* $\forall\, o \in dom(\Gamma).\ \Gamma(o) = ref C \Rightarrow h(s(o)) = (C, n, I_{(C,n)})$*, and*

*(d)* $\forall a \in \mathcal{A}. a \in dom(h) \Rightarrow h_3(a) \sim_{(s,h)} \Gamma$.

*(Definition 6.2)*

*2.* $I_{(C,n)} \sim_{(s,h)} \Gamma \Leftrightarrow^{def} \forall D. if\ C \leq D, then$

*(a)* $\Gamma((D,v)) = int \Rightarrow I_{(C,n)}(v) \in \mathbb{Z}$*, and*

*(b)* $\Gamma((D,v)) = refE \Rightarrow h(I_{(C,n)}(v)) = (E, m, I_{(E,m)})$ *and* $I_{(E,m)} \sim_{(s,h)} \Gamma$.

Now we prove the soundness of the type system.

**Lemma 1**

Typed expressions of the language J-COP do not abort (go wrong). Moreover:

(a) If $\Gamma \vDash e{:}int$ and $(s,h,L^s) \sim (\Gamma, L^t)$, then $\llbracket e \rrbracket(s,h) \in \mathbb{Z}$ .

(b) If $\Gamma \vDash e{:}ref\ C$ and $(s,h,L^s) \sim (\Gamma, L^t)$, then $h_1(\llbracket e \rrbracket(s,h)) = D$ and $D \leq C$.

**Proof**

Suppose that $e$ is an expression of the language J-COP such that $\Gamma \vDash e{:}\tau$ and $(s,h,L^s) \sim (\Gamma, L^t)$. We show that $\llbracket e \rrbracket(s,h) \neq \perp$ and we show (a) and (b) above. This is shown by induction on $\Gamma \vDash e{:}\tau$ with case analysis on the last inference rule applied. Main cases are only shown below:

Case $(o^t)$:

In this case $\Gamma(o) = \tau$. We have two subcases. In the first sub-case $\Gamma(o) = int$ which implies $s(o) \in \mathbb{Z}$ because $(s,h,L^s) \sim (\Gamma, L^t)$. In the second sub-case $\Gamma(o) = refC$ which implies $s(o) \in dom(h)$ because $(s,h,L^s) \sim (\Gamma, L^t)$. Hence in both subcases $\llbracket e \rrbracket(s,h) \neq \perp$ and clearly (a) and (b) are satisfied.

Case $(cast^t_1)$:

In this case $e = (C)e'$, $\Gamma \vDash e'{:}int$, $\Gamma \vDash$

$(C)e': int$ and $(s, h, L^s) \sim (\Gamma, L^t)$. Hence by induction hypothesis, $[\![e]\!](s,h) \in \mathbb{Z}$. Since $[\![e]\!](s,h) \notin \mathcal{A}$ , By $(cast_2^s)$, $[\![(C)e']\!](s,h) = [\![e']\!](s,h) \in \mathbb{Z}$. This completes the proof for this case.

Case $(cast_2^t)$:

In this case $e = (C)e'$, $\Gamma \vDash e': ref\ D$, $\Gamma \vDash (C)e': ref C, D \le C$, and $(s,h,L^s) \sim (\Gamma, L^t)$. Hence by induction hypothesis, $h_1([\![e']\!](s,h)) = E$ and $E \le D$ implying $E \le C$ . By $(cast_2^s)$ , $[\![(C)e']\!](s,h) = [\![e']\!](s,h)$. Hence $h_1([\![(C)e']\!](s,h)) = h_1([\![e']\!](s,h)) =$ E and $E \le C$. This completes the proof for this case.

Case $(e.v^t)$:

In this case $e = e'.v$, $\Gamma \vDash e': ref\ C$ , $class(C,v) = D$, $\Gamma((D,v)) = \tau$, and $(s,h,L^s) \sim (\Gamma, L^t)$. Hence by induction hypothesis, $h_1([\![e']\!](s,h)) = E$ and $E \le C$. Hence $E \le D$ because $C \le$ D. We also have $I = h_3([\![e']\!](s,h))$ and $v \in dom(I)$ because $class(C,v) = D$. Hence by $(inst_1^s)$, $[\![e'.v]\!](s,h) = I(v) \neq \perp$. Now $I \sim_{(s,h)} \Gamma$ because $(s,h,L^s) \sim (\Gamma, L^t)$. Hence

1. $\Gamma((D,v)) = int \Rightarrow I(v) \in \mathbb{Z}$, and
2. $\Gamma((D,v)) = ref\ \ E \Rightarrow h(I(v)) = (E, m, I_{(E,m)})$ and $I_{(E,m)} \sim_{(s,h)} \Gamma$. This completes the proof for this case.

The proof of the following lemma is similar to that of the previous one.

**Lemma 2**

Typed Boolean expressions of the language J-COP do not abort (go wrong).

$$\frac{le = \epsilon}{layer(le, Ls) = Ls}(lyr_1) \qquad \frac{le = with\ l \quad l \notin Ls}{layer(le, Ls) = [\,l \mid Ls\,]}(lyr_2) \qquad \frac{le = with\ l\ l \in Ls}{layer(le, Ls) = Ls}(lyr_3)$$

$$\frac{le = without\ l \quad remove(Ls, l) = Ls'}{layer(le, Ls) = Ls'}(lyr_4) \qquad \frac{le = le_1 le_2 \quad layer(le_1, Ls) = Ls'' \quad layer(le_2, Ls'') = Ls'}{layer(le, Ls) = Ls'}(lyr_5)$$

$$\frac{x \in IVar_C}{class(C,x) = C}(class_1) \qquad \frac{x \notin IVar_C \quad D \ll C \quad class(D,x) = E}{class(C,x) = E}(class_2)$$

$$\frac{f \in F_C}{super(C,f) = C}(super_1) \qquad \frac{f \notin F_C \quad D \ll C \quad super(D,f) = E}{super(C,f) = E}(super_2)$$

$$\frac{L_C(l) = (g,_,_,_) \qquad g \neq f}{lyrfun(C,f,l,L) = L}(lyrfun_1) \qquad \frac{L_C(l) = (f,_,_,_) \qquad g \neq f}{lyrfun(C,f,l,L) = [L \mid l]}(lyrfun_2)$$

$$\frac{\begin{array}{cc} dom(L_c) = \{l_1, \dots, l_k\} & L_1 = [] \\ lyrfun(C,f,l_i,L_i) = L_{i+1} & L' = L_{k+1} \ \cap L \end{array}}{clslyrs(C,f,L) = L'}(clslyrs)$$

**Figure 3.** Inference rules of necessary functions for semantics

$$[\![n]\!](s,h) = n \qquad [\![this]\!](s,h) = s\ (this) \qquad [\![o]\!](s,h) = s\ (o) \qquad [\![true]\!](s,h) = true$$

$$[\![false]\!](s,h) = false \qquad [\![e_1\ i_{0p}\ e_2]\!](s,h) = \begin{cases} [\![e_1]\!](s,h) i_{0p}\ [\![e_2]\!](s,h) & if\ [\![e_1]\!](s,h) i_{0p}\ [\![e_2]\!](s,h) \in \mathbb{Z}, \\ \perp & otherwise. \end{cases}$$

$$[\![e_1\ c_{0p}\ e_2]\!](s,h) = \begin{cases} [\![e_1]\!](s,h) c_{0p}\ [\![e_2]\!](s,h) & if\ [\![e_1]\!](s,h), [\![e_2]\!](s,h) \in \mathbb{Z}, \\ \perp & otherwise. \end{cases}$$

$$[\![b_1\ b_{0p}\ b_2]\!](s,h) = \begin{cases} [\![e_1]\!](s,h) b_{0p}\ [\![e_2]\!](s,h) & if\ [\![e_1]\!](s,h), [\![e_2]\!](s,h) \in \{true, false\}, \\ \perp & otherwise. \end{cases}$$

$$\frac{\begin{array}{c} [\![e]\!](s,h) \in dom(h) \\ h_1([\![e]\!](s,h)) = D \qquad not(D \le C) \end{array}}{[\![C(e)]\!]\ (s,h) = \perp}(cast_1^s) \qquad \frac{prescondition\ of\ \ (cast_1^s) is\ not\ satisfied}{[\![C(e)]\!]\ (s,h) = [\![e]\!](s,h)}(cast_2^s)$$

$$\frac{\begin{array}{c} class(h_1([\![e]\!](s,h)), v) = D \\ I = h_3([\![e]\!](s,h)) \qquad v \in dom(I) \end{array}}{[\![e.v]\!]\ (s,h) = I(v)}(inst_1^s) \qquad \frac{prescondition\ of\ \ (inst_1^s)\ is\ not\ satisfied}{[\![e.v]\!]\ (s,h) = \perp}(inst_2^s)$$

**Figure 4.** Semantics of J-COP expressions

$$\frac{\begin{array}{c} class(h_1(\llbracket e_1 \rrbracket(s,h)),v) = D \\ I = h_3(\llbracket e_1 \rrbracket(s,h)) \qquad I' = I[v \mapsto \llbracket e_2 \rrbracket(s,h)] \end{array}}{\begin{array}{c} e_1.v \coloneqq e_2 : (s,h,L^s) \to (s, h[\llbracket e_1 \rrbracket(s,h) \mapsto \\ (h_1(\llbracket e_1 \rrbracket(s,h)), h_2(\llbracket e_1 \rrbracket(s,h)), I')], L^s \ ) \end{array}} \ (\coloneqq^s_e)$$

$$\frac{\begin{array}{c} F_{h_1(\llbracket e_2 \rrbracket(s,h))}(f) = (p_f, S_f, e_f) \\ S_f : (s[this \mapsto s(o_2), p_f \mapsto \llbracket e \rrbracket(s,h)], h, L^s) \to (s', h', L^{s'}) \end{array}}{o_1 \coloneqq o_2.f(e) : (s,h,L^s) \to (s'[o_1 \mapsto \llbracket e_f \rrbracket(s',h')], h', L^{s'})} \ (\coloneqq^s_{o.f})$$

$$\frac{\begin{array}{c} s_1 = s \qquad h_1 = h \qquad layer(le, L^s) = L^s_1 \\ [l_1 \dots l_m] \subseteq L^s_1 \ such\ that\ \forall\, 1 \le i \le m \ \left(L_{h_1(\llbracket o_2 \rrbracket\,(s,h))}(l_i) = (f, S_i, e_i, p_i)\right) \\ S_1 : (s_1[this \mapsto s_1(o_2), p_1 \mapsto \llbracket e \rrbracket(s_1,h_1)], h_1, L^s_1) \to (s_2, h_2, L^s_2) \\ \forall\, i > 1.\ S_i : (s_i[o_1 \mapsto \llbracket e_{i-1} \rrbracket(s_i,h_i), this \mapsto s_i(o_2), p_i \mapsto \llbracket e \rrbracket(s_i,h_i)], h_i, L^s_i) \to a\ (s_{i+1}, h_{i+1}, L^s_{i+1}) \end{array}}{o_1 \coloneqq le\ o_2.f(e) : (s,h,L^s) \to (s_{m+1}[o_1 \mapsto \llbracket e_m \rrbracket(s_{m+1},h_{m+1})], h_{m+1}, L^s_{m+1})} \ (\coloneqq^s_{l.o.f})$$

$$\frac{\begin{array}{c} E\ is\ the\ direct\ superclass\ of\ \ h_1(s(this)) \\ super(E,f) = D \qquad F_D(f) = (p_f, S_f, e_f) \\ S_f : (s[p_f \mapsto \llbracket e \rrbracket\ (s,h)], h, L^s) \to (s', h', L^{s'}) \end{array}}{o_1 \coloneqq super.f(e) : (s,h,L^s) \to (s'[o_1 \mapsto \llbracket e_f \rrbracket\ (s',h')], h', L^{s'})} \ (sup^s)$$

$$\frac{a \in \mathcal{A} \setminus dom(h) \qquad n\ is\ fresh}{o \coloneqq new\ C : (s,h,L^s) \to (s[o \mapsto a], h[a \mapsto (C, n, \{(v,\bot) \mid v \in IVar_C\})], L^s)} \ (new^s)$$

$$\frac{\begin{array}{c} (s_1, h_1, L_1) = (s,h,L^s) \qquad clslyrs(h_1(s(o_2), f, L^s) = [l_1 \dots l_m] \\ \forall\, 1 \le i \le m\ (L_C(l_i)) = (f, p_i, S_i, e_i)) \\ S_1 : (s_1[this \mapsto s(o_2), p_1 \mapsto \llbracket e \rrbracket(s_1,h_1)], h_1, L^s_1) \to (s_2, h_2, L^s_2) \\ \forall\, i > 1. S_i : (s_i[o_1 \mapsto \llbracket e_{i-1} \rrbracket(s_i,h_i), this \mapsto s(o_2), p_i \mapsto \llbracket e \rrbracket(s_i,h_i)], h_i, L^s_i) \to (s_{i+1}, h_{i+1}, L^s_{i+1}) \end{array}}{o_1 \coloneqq proceed\ o_2.f(e) : (s,h,L^s) \to (s_{m+1}[o_1 \mapsto \llbracket e_m \rrbracket(s_{m+1},h_{m+1})], h_{m+1}, L^s_{m+1})} \ (pro^s)$$

$$\frac{\begin{array}{c} \left(\llbracket b \rrbracket(s,h) = true \wedge S_t : (s,h,L^s) \to (s',h',L^{s'})\right) \vee \\ \left(\llbracket b \rrbracket(s,h) = false \wedge S_f : (s,h,L^s) \to (s',h',L^{s'})\right) \end{array}}{if\ b\ then\ S_t else\ S_f : (s,h,L^s) \to (s',h',L^{s'})} \ (if^s) \qquad \frac{\llbracket b \rrbracket(s,h) = false}{while\ b\ \ do\ S_t : (s,h,L^s) \to (s,h,L^s)} \ (while^s_1)$$

$$\frac{\begin{array}{c} S_1 : (s,h,L^s) \to (s'',h'',L^{s''}) \\ S_2 : (s'',h'',L^{s''}) \to (s',h',L^{s'}) \end{array}}{S_1; S_2 : (s,h,L^s) \to (s',h',L^{s'})} \ (Seq^s) \qquad \frac{\begin{array}{c} \llbracket b \rrbracket(s,h) = true \\ S_t : (s,h,L^s) \to (s'',h'',L^{s''}) \\ while\ b\ \ do\ S_t : (s'',h'',L^{s''}) \to (s',h',L^{s'}) \end{array}}{while\ b\ \ do\ S_t : (s,h,L^s) \to (s',h',L^{s'})} \ (while^s_2)$$

**Figure 5.** Inference rules of the operational semantics for J-COP constructs

**Theorem 1**

Well-formed statements of the language J-COP do not abort (go wrong).

**Proof**

Suppose that $S$ is a statement of the J-COP language. Suppose that the maps $F_C$ and $L_C$ and the relation $\le$ describing the classes used in $S$ are given along with $S$. Suppose also that $(\Gamma, L^t) \vDash S : WF$ and $(s,h,L^s) \sim (\Gamma, L^t)$. We show that if $S$ does not contain infinite loop then $\neg(S : (s,h,L^s) \to abort)$, i.e. there is a state $(s', h', L^{s'})$ such that $S : (s,h,L^s) \to (s',h',L^{s'})$. Moreover we show that in this case $(s', h', L^{s'}) \sim (\Gamma', L^t)$ where $\Gamma' = \Gamma](dom(\Gamma) \setminus \{this, p\})$. This is shown by induction on $(\Gamma, L^t) \vDash S : WF$ with case analysis on the last type rule applied. Outlines of main cases are shown below.

Case $(\coloneqq^t_e)$

$$\frac{}{\Gamma \vDash n: int}\,(int^t) \qquad \frac{\Gamma(this) = \tau}{\Gamma \vDash this: \tau}\,(this^t) \qquad \frac{}{\Gamma \vDash true: bool}\,(true^t) \qquad \frac{\Gamma(o) = \tau}{\Gamma \vDash o: \tau}\,(o^t) \qquad \frac{\Gamma \vDash e_1, e_2: int}{\Gamma \vDash e_1\, c_{op}\, e_2: bool}\,(c^t_{op})$$

$$\frac{\Gamma \vDash b_1, b_2: bool}{\Gamma \vDash b_1\, b_{op}\, b_2: bool}\,(b^t_{op}) \qquad \frac{\Gamma \vDash e: int}{\Gamma \vDash: (C)e: int}\,(cast^t_1) \qquad \frac{\Gamma \vDash e: ref\ D \quad D \le C}{\Gamma \vDash: (C)e: ref\ C}\,(cast^t_2)$$

$$\frac{\Gamma \vDash e_1, e_2: int}{\Gamma \vDash e_1\, i_{op}\, e_2: int}\,(i^t_{op}) \qquad \frac{\Gamma \vDash e: ref\ C \quad class(C, v) = D \quad \Gamma\big((D, v)\big) = \tau}{\Gamma \vDash e.v: \tau}\,(e.v^t)$$

$$\frac{\begin{array}{c}\Gamma \vDash o_2: ref\ C \quad \Gamma \vDash o_1: \tau_2' \quad \tau_1' \le \tau_1 \\ (\Gamma, L^t) \vDash (C, f): \tau_1 \to \tau_2 \quad \Gamma \vDash e: \tau_1' \quad \tau_2 \le \tau_2'\end{array}}{(\Gamma, L^t) \vDash o_1 := o_2.f(e): WF}\,(:=^t_{o.f})$$

$$\frac{\begin{array}{c}F_C(f) = (p, S, e') \qquad (\Gamma[p \mapsto \tau_1, this \mapsto ref\ C], L^t) \vDash S: WF \\ \Gamma \vDash p: \tau_1 \qquad \Gamma[p \mapsto \tau_1, this \mapsto ref\ C] \vDash e': \tau_2 \\ \forall D.(C \ll D)\, it\ is\ true\ that\ \big((\Gamma, L^t) \vDash (D, f): \iota_1 \to \iota_2\big) \Rightarrow (\iota_1 = \tau_1\ and\ \tau_2 \le \iota_2)\end{array}}{(\Gamma, L^t) \vDash (C, f): \tau_1 \to \tau_2}\,(C.f^t)$$

$$\frac{\Gamma \vDash e_1.v: \tau_1 \quad \Gamma \vDash e_2: \tau_2 \quad \tau_2 \le \tau_1}{(\Gamma, L^t) \vDash e_1.v := e_2: WF}\,(:=^t_e) \qquad \frac{\begin{array}{c}L_C(l) = (f, p_l, S_l, e_l) \quad \Gamma \vDash p_l: \tau_1 \\ (\Gamma[this \mapsto ref\ C,\ p \mapsto \tau_1], L^t) \vDash S_1: WF \\ \Gamma[this \mapsto ref\ C, p \mapsto \tau_1] \vDash e: \iota_l\end{array}}{(\Gamma, L^t) \vDash (C, f, l): \tau_1 \to \iota_l}\,(C.l.f^t)$$

$$\frac{\begin{array}{c}\exists \tau_1, \tau_2. \forall\, l \in L^t.\ if\ L_C(l) = (f, _, _, _),\ then\ (\Gamma, L^t) \vDash (C, f, l): \tau_1 \to \iota_l\ and\ \tau_2 \le \iota_l \\ \Gamma \vDash o_1: \tau_2' \quad \tau_1' \le \tau_1 \quad \tau_2 \le \tau_2' \quad layer(le, L^t) = L^t \quad \Gamma \vDash o_2: ref\ C \quad \Gamma \vDash e: \tau'_1\end{array}}{(\Gamma, L^t) \vDash o_1 := le\ o_2.f(e): WF}\,(:=^t_{l.o.f})$$

$$\frac{\begin{array}{c}\Gamma \vDash this: ref\ C \quad (\Gamma, L^t) \vDash (D, f): \tau_1 \to \tau_2 \\ C \ll D \quad \Gamma \vDash o_1: \tau_2' \quad \Gamma \vDash e: \tau_1' \quad \tau_1' \le \tau_1 \quad \tau_2 \le \tau_2'\end{array}}{(\Gamma, L^t) \vDash o_1 := super.f(e): WF}\,(sup^t)$$

$$\frac{\begin{array}{c}\forall\, l \in L.\, if\ (L_C(l) = (f, p, S, e)\ and\ (\Gamma, L^t) \vDash (C, f, l): \iota_1 \to \iota_2, then\ (\iota_1 = \tau_1\ and\ \tau_2 \le \iota_2) \\ \Gamma \vDash o_2: ref\ C \quad \Gamma \vDash e: \tau_1' \quad \tau_1' \le \tau_1 \quad \Gamma \vDash o_1: \tau_2' \quad \tau_2 \le \tau_2'\end{array}}{(\Gamma, L^t) \vDash o_1 := proceed\ o_2.f(e): WF}\,(pro^t)$$

$$\frac{\Gamma \vDash o: \tau \quad \tau \le ref\ C}{(\Gamma, L^t) \vDash o := new\ C: WF}\,(new^t) \qquad \frac{\Gamma \vDash b: bool \quad (\Gamma, L^t) \vDash S_t, S_f: WF}{(\Gamma, L^t) \vDash if\ \ b\ then\ S_t\ else\ S_f: WF}\,(if^t)$$

$$\frac{\Gamma \vDash \mathrm{b: bool} \quad (\Gamma, L^t) \vDash S_t: WF}{(\Gamma, L^t) \vDash \mathrm{while\ b\ do}\ S_t: WF}\,(while^t) \qquad \frac{(\Gamma, L^t) \vDash S_1: WF \quad (\Gamma, L^t) \vDash S_2: WF}{\Gamma \vDash S_1; S_2: WF}\,(seq^t)$$

**Figure 6.** Type system for J-COP constructs**.**

In this case:

a. $S = e_1.v := e_2$.

b. $\Gamma \vDash e_1.v: \tau_1, \Gamma \vDash e_2: \tau_2, and\ \tau_2 \le \tau_1$.

c. By the rule $(e.v^t)$, $\Gamma \vDash e_1.v: \tau_1 \Rightarrow \Gamma \vDash e_1: ref\ C,\ class(C, v) = E$, and $\Gamma\big((E, v)\big) = \tau_1$.

By Lemma 1, $[\![e_1]\!](s, h) \in dom(h)$ and $h_1\big([\![e_1]\!](s, h)\big) \le C$. Hence there is a class $D$ such that $class(h_1([\![e_1]\!](s, h)), v) = D$ because there is a class $E$ such that $class(C, v) = E$ . Also by Lemma 1, $[\![e_2]\!](s, h) \neq \perp$. Hence the state $(s', h', L^{s\prime}) = (s, h[[\![e_1]\!](s, h) \mapsto \big(h_1\big([\![e_1]\!](s, h)\big), h_2\big([\![e_1]\!](s, h)\big), I'\big), L^s)$ is defined and the statement does not abort. Clearly $(s', h', L^{s\prime}) \sim (\Gamma', L^t)$.

Case $(:=^t_{o.f})$

In this case:

a. $S = o_1 \coloneqq o_2.f(e)$.
b. $\Gamma \vDash o_2{:}ref\ C, \Gamma \vDash o_1{:}\tau_2', \Gamma \vDash e{:}\tau_1', \tau_1' \leq \tau_1$, and $\tau_2 \leq \tau_2'$.
c. $(f) = (p, S, e')$ and $\Gamma \vDash p{:}\tau_1$.
d. $(\Gamma[p \mapsto \tau_1, this \mapsto ref\ C], L^t) \vDash S{:}WF$ and $\Gamma[p \mapsto \tau_1, this \mapsto ref\ C] \vDash e'{:}\tau_2$.
e. $\forall D.(C \ll D)$ *it is true that:*
$$((\Gamma, L^t) \vDash (D, f){:}\iota_1 \to \iota_2) \Rightarrow (\iota_1 = \tau_1 and\ \tau_2 \leq \iota_2).$$

By Lemma 1, $[\![o_2]\!](s,h) \in dom(h)$ and $h_1([\![o_2]\!](s,h)) \leq C$. Since $F_C(f) = (p, S, e')$, $F_{h_1([\![o_2]\!](s,h))}(f) = (p, S, e')$. Now we notice that $(s[this \mapsto s(o_2), p \mapsto [\![e]\!](s,h)], h, L^s) \sim (\Gamma[p \mapsto \tau_1, this \mapsto ref\ C], L^t)$. This is so because:

1. $h(s(this)) = h(s(o_2)) = (C, n, I_{(C,n)})$ because $\Gamma \vDash o_2{:}ref\ C$, and
2. $s(p) = s([\![e]\!](s,h))$ and $\Gamma \vDash e{:}\tau_1' \leq \tau_1$.

Therefore by induction hypothesis there is a state $(s', h', L^{s\prime})$ such that $(s', h', L^{s\prime}) \sim (\Gamma[p \mapsto \tau_1, this \mapsto ref\ C], L^t)$ and
$$S : (s[this \mapsto s(o_2), p \mapsto [\![e]\!](s,h)], h, L^s) \to (s', h', L^{s\prime})$$
because
$$(\Gamma[p \mapsto \tau_1, this \mapsto ref\ C], L^t) \vDash S{:}WF.$$
This implies $[\![e']\!](s', h') \neq \perp$ because
$$\Gamma[p \mapsto \tau_1, this \mapsto ref\ C] \vDash e'{:}\tau_2.$$
Hence the state $(s[o_1 \mapsto [\![e']\!](s', h')], h', L^{s\prime})$ is defined and satisfies $(s'[o_1 \mapsto [\![e']\!](s', h')], h', L^{s\prime}) \sim (\Gamma', L^t)$. This completes the proof of this case.

Case $(\coloneqq^t_{l.o.f})$

In this case

a. $S = o_1 \coloneqq le\ o_2.f(e)$.
b. $layer(le, L^s) = L_1^s$ and $(s_1, h_1, L_1^s) = (s, h, L)$.
c. $[l_1 \dots l_m] \subseteq L_1^s$ such that $\forall 1 \leq i \leq m \left(L_{\{h_1([\![o_2]\!](s,h))\}(l_i)} = (f, p_i, S_i, e_i)\right)$.
d. $\exists \tau_1, \tau_2$ such that for all $l \in L^t$ *if* $L_C(l) = (f, p_l, S_l, e_l)$, *then* $\Gamma \vDash p_l{:}\tau_1, \Gamma[p \mapsto \tau_1, this \mapsto ref\ C] \vDash S_l{:}WF$, *and* $\Gamma[p \mapsto \tau_1, this \mapsto ref\ C] \vDash e_l{:}\iota_l$.
e. $\Gamma \vDash o_2{:}ref\ C, \Gamma \vDash e{:}\tau_1', \Gamma \vDash o_1 : \tau_2', \tau_1' \leq \tau_1$, and $\tau_2 \leq \tau_2'$.

By Lemma 1, $[\![o_2]\!](s,h) \in dom(h)$ and $h_1([\![o_2]\!](s,h)) \leq C$. If the list $[L_1 \dots L_m]$ is empty then the statement $S$ does not abort and $(s', h', L^{s\prime}) = (s, h, L^s)$. For $l_1$, we have $L_{h_1([\![o_2]\!](s,h))}(l_1) = (f, p_1, S_1, e_1)$. Similarly to the previous case, we conclude:
$$(s_1[this \mapsto s_1(o_2), p_1 \mapsto [\![e]\!](s_1, h_1), h_1, L_1^s) \sim (\Gamma[p_1 \mapsto \tau_1, this \mapsto ref\ C], L^t).$$
Then by induction hypothesis there exists a state $(s_2, h_2, L_2^s)$ such that) $(s_2, h_2, L_2^s) \sim (\Gamma', L^t)$ and
$$S_1{:}(s_1[this \mapsto s_1(o_2), p_1 \mapsto [\![e]\!](s_1, h_1)], h_1, L_1) \to (s_2, h_2, L_2^s).$$

Now clearly,
$(s_2[o_1 \mapsto [\![e_1]\!](s_2, h_2), this \mapsto s_2(o_2), p_2 \mapsto [\![e]\!](s_2, h_2)], h_2, L_2^s) \sim (\Gamma[p_1 \mapsto \tau_1, this \mapsto ref\ C], L^t)$.
Then by induction hypothesis there exists a state $(s_3, h_3, L_3^s)$ such that) $(s_3, h_3, L_3^s) \sim (\Gamma', L^t)$ and
$$S_2{:}(s_2[o_1 \mapsto [\![e_1]\!](s_2, h_2), this \mapsto s_2(o_2), p_2 \mapsto [\![e]\!](s_2, h_2)], h_2, L_2) \to (s_3, h_3, L_3^s).$$
Therefore a simple induction on $m$ can prove that for all $i$ there exists state $(s_{i+1}, h_{i+1}, L_{i+1}^s)$ such that) $(s_{i+1}, h_{i+1}, L_{i+13}^s) \sim (\Gamma', L^t)$ and
$$\forall i\ S_i{:}(s_2[o_1 \mapsto [\![e_{i-1}]\!](s_i, h_i), this \mapsto s_i(o_2), p_i \mapsto [\![e]\!](s_i, h_i)], h_2, L_2) \to (s_{i+1}, h_{i+1}, L_{i+1}^s).$$
Hence the sate $(s_{m+1}, h_{m+1}, L_{m+1}^s)$ is defined and satisfies $(s_{m+1}, h_{m+1}, L_{m+1}^s) \sim (\Gamma', L^t)$. Now by Lemma 1, $[\![e_m]\!](s_{m+1}, h_{m+1}) \neq \perp$ and hence $(s_{m+1}[o_1 \mapsto [\![e_m]\!](s_{m+1}, h_{m+1})], h_{m+1}, L_{m+1}^s)$ is defined and satisfies $(s_{m+1}[o_1 \mapsto [\![e_m]\!](s_{m+1}, h_{m+1})], h_{m+1}, L_{m+1}^s) \sim (\Gamma', L^t)$ which completes the proof of this case.

Case $(sup^t)$:

This case is similar to the case of $(\coloneqq^t_{o.f})$.

Case $(pro^t)$:

This case is similar to the case of $(\coloneqq^t_{l.o.f})$.

## 4. Discussions

**Related work**: an operational semantics and a type system for modeling and checking context-oriented constructs are presented in (Hirschfeld, Igarashi & Masuhara, 2011). While the language model studied in (Hirschfeld et al., 2011) is functional, our model is structural. The type system presented in the current paper and that in (Hirschfeld et al., 2011) stop the command *proceed* from executing faulty procedures.

An operational semantics, that is based on delegation based calculus, is presented in (Schippers, Janssens, Haupt & Hirschfeld, 2008) for the language $c_j$, a context-oriented programming language. The research in (Clarke & Sergey, 2009) presents a syntax-based semantics for COP concepts as implemented by ContextL, ContextJ*, and other examples. This paper also introduces a type system that prevents program from getting stuck. The semantics presented of most related work uses general calculi to represent context-dependent behavior of COP programs. Our semantics, on the other hand, is built directly on an accurate memory model which adds to the clarity and soundness of our semantics and type system.

Aiming at describing behavioral variations, delta modules and layers are used by delta-oriented programming (DOP) (Schaefer, Bettini, Bono,

Damiani & Tanzarella, 2010) and feature-oriented programming (FOP) (Batory, Sarvela & Rauschmayer, 2004), respectively. In these manners, the static composition of classes with layers creates many similar software artifacts. Transitional semantics was presented for DOP in (Schaefer, Bettini & Damiani, 2011) and for FOP in (Delaware, Cook & Batory, 2009). Noticeably, in these approaches new procedures can be added by layers. This fact sophisticates any accurate semantics and type system for DOP and FOP.

**Future work**: it is intersecting to extend the language of the current paper to allow layer inheritance and layer dependency. This enables one layer to require the presence of another layer. It also enables expressing the condition that two layers cannot be active simultaneously. Another direction for a future work is to extend the language to associate candidate procedures of the command *proceed()* with priorities for execution.

**Acknowledgements:**

Foundation item: Al Imam University (IMSIU) Project (No.: 330917). Authors are grateful to Al Imam University (IMSIU), KSA for financial support to carry out this work.

**Corresponding Author:**
Dr. Mohamed A. El-Zawawy
College of Computer and Information Sciences, Al Imam Mohammad Ibn Saud Islamic University (IMSIU), Riyadh, Kingdom of Saudi Arabia.
E-mail: maelzawawy@cu.edu.eg

2/2/2013